\documentclass[fleqn,usenatbib]{mnras}
\usepackage{newtxtext,newtxmath}
\usepackage[T1]{fontenc}
\DeclareRobustCommand{\VAN}[3]{#2}
\let\VANthebibliography\thebibliography
\def\thebibliography{\DeclareRobustCommand{\VAN}[3]{##3}\VANthebibliography}
\usepackage{graphicx}	
\usepackage{amsmath}	
\usepackage{color}

\def\plotone#1{\includegraphics[width=0.5\textwidth]{#1}}
\title[$\tau$ in punctuated inflation]{Reionization optical depth and CMB-BAO tension in punctuated inflation}

\author[Zhiqi Huang]{
Zhiqi Huang$^{1,2}$\thanks{E-mail: huangzhq25@mail.sysu.edu.cn}
\\
${}^{1}$ School of Physics and Astronomy, Sun Yat-sen University, 2 Daxue Road, Tangjia, Zhuhai, 519082, China \\
${}^{2}$ CSST Science Center for the Guangdong-Hongkong-Macau Greater Bay Area, Sun Yat-sen University, Zhuhai, 519082, China 
}

\date{Accepted XXX. Received YYY; in original form ZZZ}

\pubyear{2025}

\begin{document}
\label{firstpage}
\pagerange{\pageref{firstpage}--\pageref{lastpage}}
\maketitle

\begin{abstract}
  Within the standard six-parameter Lambda cold dark matter ($\Lambda$CDM) model, a $2$-$3\sigma$ tension persists between baryon acoustic oscillation (BAO) measurements from the Dark Energy Spectroscopic Instrument (DESI) and observations of the cosmic microwave background (CMB). Although this tension has often been interpreted as evidence for dynamical dark energy or a sum of neutrino masses below the established minimum, recent studies suggest it may instead originate from an underestimation of the reionization optical depth, particularly when inferred from large-scale CMB polarization. Jhaveri et al. propose that a suppression of large-scale primordial curvature power could partially cancel the contribution of $\tau$ to the CMB low-$\ell$ polarization power spectrum, leading to a biased low $\tau$ measurement in standard analyses. In this work, we investigate whether punctuated inflation - which generates a suppression of primordial power on large scales through a transient fast-roll phase - can raise the inferred $\tau$ value and thereby reconcile the consistency between CMB and BAO. For simple models with step-like features in the inflaton potential, we find that the constraint on $\tau$ and the CMB-BAO tension remain nearly identical to those in the standard six-parameter $\Lambda$CDM model. We provide a physical explanation for this negative result.
\end{abstract}



\section{Introduction} \label{sec:intro}

\begin{figure}
  \centering
  \plotone{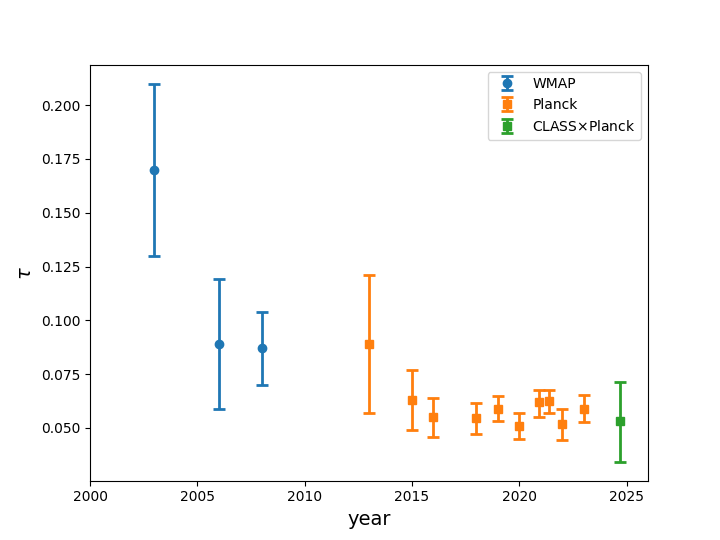}
  \caption{The evolution of CMB constraint on $\tau$ in $\Lambda$CDM cosmology. \label{fig:tauhis}}
\end{figure}

The reionization optical depth (denoted as $\tau$) is a pivotal cosmological parameter quantifying the total electron column density encountered by cosmic microwave background (CMB) photons since the epoch of reionization. Thomson scattering during reionization generates large-scale anisotropy of CMB E-mode polarization, with the optical depth affecting the amplitude of the polarization power spectrum at low multipoles $\ell\lesssim 10$~\citep{Kaplinghat03}. These low-$\ell$  polarization multipoles are difficult to measure with ground-based CMB experiments, which typically have limited sky coverage and suffer from ground and atmosphere noises at low-$\ell$. (See however \cite{CLASS25} which makes use of the cross correlation between the data from ground-based mission and space mission.) To date, the space missions Wilkinson Microwave Anisotropy Probe (WMAP) and Planck have provided the most accurate determination of $\tau$. Figure~\ref{fig:tauhis} shows the historical evolution of $\tau$ measurements under the standard Lambda cold dark matter ($\Lambda$CDM) cosmology~\citep{WMAP03Params, WMAP07Params, WMAP09Params, Planck13Params, Planck15Params, Planck16Tau, Planck18Params, Planck19Tau, Planck20NPIPE, Planck21ReLike, deBelsunce21, Planck22PR4, Planck23PR4, CLASS25}. Currently the best constraint on $\tau$ is mainly from the Planck 2018 low-$\ell$ EE data. Since 2018, results derived from the same Planck 2018 data exhibit $\sim 0.5\sigma$ fluctuations in the mean $\tau$ value. These reflect ongoing efforts to build a more complete model of instrumental and foreground uncertainties~\citep{SRoll2, BeyondPlanckI}.

On small scales, the integrated contribution from line-of-sight reionization bubbles results in an overall $e^{-2\tau}$ suppression of the CMB power spectra. Thus, at high multipoles ($\ell\gtrsim 10$), the observed power spectra tightly constrain the parameter combination $A_s e^{-2\tau}$, where $A_s$ is the amplitude of the primordial scalar perturbation power spectrum. On the other hand, CMB lensing directly probes $A_s$ and the matter abundance parameter $\Omega_m$. Therefore, under $\Lambda$CDM cosmology - where $\Omega_m$ is well determined - $\tau$ can also be constrained by CMB data excluding low-$\ell$ polarization (CMB-no-lowP). Low-redshift geometric probes such as baryon acoustic oscillations (BAO) may further tighten the constraint on $\Omega_m$ and hence improves the CMB-no-lowP constraint on $\tau$. With Planck 2018 PR3 low-$\ell$ TT and high-$\ell$ plik TTTEEE~\citep{Planck18Params}, and a joint lensing likelihood from Planck 2018 PR4~\citep{Planck22PR4lensing}, ACT DR6~\citep{ACTDR6lensing}, and SPT-3G MUSE~\citep{SPT3Glensing}, recent work \cite{Jhaveri25} (hereafter J25) found a CMB-no-lowP constraint $\tau = 0.080\pm 0.016$, which is about $1.5\sigma$ higher than the full Planck 2018 PR3 result $\tau = 0.0544\pm 0.0073$. Adding the measurement of baron acoustic oscillations (BAO) from Dark Energy Spectroscopic Instrument (DESI)~\citep{DESIDR2} further raises $\tau$ to $0.091\pm 0.011$, which is $2.8\sigma$ higher than the full CMB constraint. Similar results has been found in independent studies~\citep{Sailer25, Allali25a, Allali25b}.

The $\tau$ tension implies some inconsistency between DESI BAO and CMB under $\Lambda$CDM, which was first revealed in the 2024 DESI release from different perspectives~\citep{DESIDR1}. Within the $\Lambda$CDM framework, BAO data provide constraints on $\Omega_m$ and the combination $hr_d$ (with $r_d$ being the sound horizon at the baryon drag epoch and $h = \frac{H_0}{100\,\mathrm{km/s/Mpc}}$ the reduced Hubble constant). The tension between DESI BAO and CMB results is therefore most apparent in the $\Omega_m$–$hr_d$ plane, where the 95\% confidence regions favored by the two data sets show little overlap~\citep{DESIDR2}. The latest release of CMB data from South Pole Telescope further shrinks the CMB contour and increase the $\Omega_m$-$hr_d$ tension to $\gtrsim 2.5\sigma$~\citep{SPT25}.

The tension between CMB and DESI BAO measurements is often interpreted as evidence for dynamical dark energy, particularly in joint analyses incorporating Type Ia supernovae~\citep{DESIDR1, DESIDR2, DESI_DE, DESI_DDE, Giare25a, Giare25b, Lee25, Huang24}. However, the preferred parameter space for such dark energy typically requires phantom crossing—a behavior likely unphysical for a single, non-interacting fluid~\citep{Tsujikawa25, Lewis25}. An alternative solution, arguably even less physical, is to allow the sum of neutrino masses to fall significantly below the minimum value ($0.06\,\mathrm{eV}$) established by neutrino oscillation experiments~\citep{Esteban24, DESIDR1, DESIDR2, Craig24, Elbers25a, Green25, Lynch25, Chebat25}. Beyond dynamical dark energy and sub-minimal neutrino mass, numerous extended cosmological models have been explored~\citep{Chen25, Pang25, Jhaveri25, Sailer25, Ferreira25, German25, Choudhury24, Choudhury25}. It has been demonstrated that excluding the CMB low-$\ell$ polarization data (specifically the Planck low-$\ell$ EE likelihood) or fixing the optical depth $\tau = 0.09$ significantly alleviates the CMB-BAO tension~\citep{Jhaveri25,Sailer25}. However,  since at present there is no evidence for large systematics in the Planck low-$\ell$ data, neither excluding the Planck low-$\ell$ EE likelihood nor fixing $\tau=0.09$ is justified under the six-parameter $\Lambda$CDM model. These studies, however, points to a novel possibility of resolving the CMB-BAO tension by elevating $\tau$ within beyond-$\Lambda$CDM models. Because the correlation between $\tau$, $\Omega_m$, and $hr_d$ is model dependent, this solution remains a conjecture before a concrete model is demonstrated to work.

Given the strong $\tau$-$A_s$ degeneracy, it is natural to seek a higher $\tau$ value by extending the power-law model for the primordial scalar spectrum. Along these lines, J25 explored a specific phenomenological extension invoking an exponential cutoff on large scales. The suppression of large-scale power cancels part of the $\tau$ contribution to the CMB EE power spectrum, leading to an underestimation of $\tau$ if the data is interpreted under the six-parameter $\Lambda$CDM model. However, J25 find the reduction of large scale power in the exponential-cutoff form has quite limited impact on $\tau$ and does not reconcile the consistency between DESI BAO and CMB.

Physically, suppression of power on the largest scales can be achieved in punctuated inflation models, which feature a transient fast-roll phase during inflation. To localize this suppression specifically to the lowest CMB multipoles (e.g., $\ell\lesssim 10$), moderate fine-tuning of either the initial field conditions or the functional form of the inflaton potential is required~\citep{Contaldi03}. A transient fast-roll phase not only introduces the dominant suppression feature but also typically generates subdominant oscillatory features that are not captured by the exponential cutoff model studied in J25. Crucially, when the suppression is tuned to low multipoles, these oscillatory features tend to manifest at higher multipoles—where CMB power spectra are measured with greater precision. Consequently, the features missed in J25 may have a significant impact on parameter inference. To assess this potential impact, the present work aims to investigate the possibility of alleviating the CMB-BAO tension with the punctuated inflation model. 

This paper is organized as follows. Section~\ref{sec:model} specifies the model under investigation.  In Section~\ref{sec:results} we use CMB and DESI BAO data to constrain cosmological parameters. Section~\ref{sec:conc} discusses and concludes. Throughout the paper we use natural units $c=\hbar=1$ and the reduced Planck mass $M_p \equiv \frac{1}{\sqrt{8\pi G_N}}$, where $G_N$ is the Newton's gravitational constant. The scale factor is denoted as $a$, and $H$ represents the Hubble parameter. An over-dot and a prime denote derivatives with respect to the cosmological time $t$ and the conformal time $\tau = \int\frac{\mathrm{d}t}{a}$, respectively. The conformal Hubble parameter is defined as $\mathcal{H} = \frac{a^\prime}{a} = aH = \dot{a}$. 

\section{Model} \label{sec:model}

The punctuated inflation model is characterized by a fast-roll phase sandwiched between slow-roll phases. Such cases can arise in double inflationary scenarios~\citep{Polarski95, Adams97, Tsujikawa03, Schettler16}, or some glitches in single field scenario~\citep{Jain09, Jain10, Goswami13, Qureshi17, Ragavendra21, Ragavendra22}. In the present work we focus on the simplest single-field case where the transient fast-roll phase is induced by a step-like feature in the inflaton potential, $V(\phi)$. 

We start with a slow-roll inflaton potential $V_{\rm SR}(\phi)$ without the step-like feature. The slow-roll phase is parameterized with the scalar amplitude $A_s$, scalar tilt $n_s$ and tensor-to-scalar ratio $r$. These parameters are defined at a pivot scale $k_p = 0.05\,\mathrm{Mpc}^{-1}$. The primordial scalar and tensor power spectra in the slow-roll approximation are given by $\mathcal{P}_S(k) = A_s\left(k/k_p\right)^{n_s-1}$ and $\mathcal{P}_T(k) = rA_s\left(k/k_p\right)^{-r/8}$, respectively.

Truncated at the second derivative of the potential, the slow-roll parameters are~\citep{Stewart93, Liddle94a, Liddle94b}
\begin{eqnarray}
  \epsilon_V &=& \frac{r}{16}, \\
  \delta &=& \frac{1-n_s}{2}, \\
  \eta_V &=& 3\epsilon_V - \delta, \\
  \epsilon_H &=& \epsilon_V(1-\frac{4}{3}\epsilon_V + \frac{2}{3}\eta_V).
\end{eqnarray}
At the pivot time, which is defined as the time when the pivot scale exits the horizon ($k_p=\mathcal{H}$), the Hubble parameter (up to the linear order of slow-roll parameters) is
\begin{equation}
  H_{\rm pivot} = 2\pi M_p \sqrt{\frac{2\epsilon_H A_s}{1-2\epsilon_H + 2(2-\ln 2 - \gamma)\delta}},
\end{equation}
where $\gamma\approx 0.5772$ is the Euler-Mascheroni constant.
Since single-field model is invariant under a field translation/reflection, we may assume the inflaton value $\phi = 0$ and the field velocity $\dot\phi < 0$ at the pivot time. Consequently, the slow-roll potential can be locally approximated with
\begin{equation}
  V_{\rm SR}(\phi) \approx (3-\epsilon_H)H_{\rm pivot}^2M_p^2 \left[ 1 + \sqrt{2\epsilon_V}\frac{\phi}{M_p} + \frac{\eta_V}{2}\left(\frac{\phi}{M_p}\right)^2\right].
\end{equation}
The  background kinetic energy and the field velocity at the pivot time are
\begin{equation}
  K_{\rm pivot} = \epsilon_H H_{\rm pivot}^2M_p^2,
\end{equation}
and
\begin{equation}
  {\dot\phi}_{\rm pivot} = -\sqrt{2K_{\rm pivot}}.
\end{equation}
To generate a fast-roll feature at large scales $\sim k_f\ll k_p$, we add the step-like feature at 
\begin{equation}
  \phi_f \equiv \frac{\dot\phi_{\rm pivot}}{H_{\rm pivot}}\ln \frac{k_f}{k_p}.
\end{equation}
The logarithmic width of the fast-roll feature in wavenumber space, which we denote as $\delta\ln k_f$, can be converted to the width in field space,
\begin{equation}
  \delta\phi_f \equiv \left\vert\frac{\dot\phi_{\rm pivot}}{H_{\rm pivot}}\right\vert \delta\ln k_f.
\end{equation}
Finally, we introduce an amplitude parameter $\lambda\ge 0$ to measure the energy gap in unit of slow-roll kinetic energy and per $\ln k$. The inflaton potential is parameterized with
\begin{equation}
  V(\phi) = V_{\rm SR}(\phi) +  \lambda K_{\rm pivot}\;\delta\ln k_f\; \mathrm{erf}\left(\frac{\phi-\phi_f}{\delta\phi_f}\right), \label{eq:erf}
\end{equation}
which is referred to as the $\mathrm{erf}$ model, or
\begin{equation}
  V(\phi) = V_{\rm SR}(\phi) +  \lambda K_{\rm pivot}\;\delta\ln k_f\; \mathrm{tanh}\left(\frac{\phi-\phi_f}{\delta\phi_f}\right), \label{eq:tanh}
\end{equation}
which is named the $\tanh$ model.
In the limit of $\lambda = 0$, both models reduce to the standard $\Lambda$CDM cosmology but with an additional parameter $r$ that permits a non-zero tensor contribution. We designate this $\lambda=0$ case as the no-step model ($\Lambda$CDM+$r$).

We summarize the cosmological parameters for the punctuated inflation model in Table~\ref{tab:params}, along with their priors for Markov Chain Monte Carlo (MCMC) simulations and their default values that we will implicitly use for demonstration of concrete examples. In the results we will also present a few derived parameters including $\sigma_8$ (root mean square fluctuation of matter in a sphere with radius $8h^{-1}\mathrm{Mpc}$) and $S_8=\sigma_8\left(\frac{\Omega_m}{0.3}\right)^{1/2}$.
\begin{table*}
  \caption{Cosmological parameters \label{tab:params} }
  \begin{tabular}{llll}
    \hline
    \hline
    parameter & definition & uniform prior & default value \\
    \hline
    $\Omega_b h^2$ & baryon density & $[0.005, 0.1]$ & $0.022383$ \\
    $\Omega_c h^2$ & CDM density & $[0.001, 0.99]$ & $0.12011$ \\
    $100\theta_{\rm MC}$ & angular extension of sound horizon at recombination & $[0.5, 10]$ & $1.040909$ \\
    $\tau$ & reionization optical depth & $[0.01, 0.2]$ & $0.0543$ \\
    $\ln\left[10^{10}A_s\right]$ & logarithmic amplitude of scalar power &  $[1.61, 3.91]$ & $2.9362+2\tau$ \\
    $n_s$ & scalar tilt & $[0.92, 1.05]$ & $0.96605$ \\
    $r$ & tensor-to-scalar ratio  & $[0, 0.2]$ & $0.01$ \\
    $\lambda$ & energy gap per $\ln k$ & $[0, 10]$ & $0$ \\
    $\ln\frac{k_f}{k_p}$ & location of step-like feature & $[-6.908, -3.912]$  & $-5.4$ \\
    $\delta\ln k_f$ & width of step-like feature & $[0.1, 2]$ & $1$ \\
    \hline
  \end{tabular}
\end{table*}

\section{Method and Results} \label{sec:results}

We solve the background equations
\begin{eqnarray}
  \ddot\phi + 3H\dot\phi + \frac{dV}{d\phi} &=& 0, \\
  \dot{H} &=& -\frac{\dot\phi^2}{2M_p^2},
\end{eqnarray}
with the initial conditions determined by slow-roll conditions at $\phi - \phi_f \gg \delta\phi_f$ (the first slow-roll phase).

The conservation equation
\begin{equation}
  3H^2M_p^2 = V(\phi) + \frac{1}{2}\dot\phi^2 \label{eq:bg_cons}
\end{equation}
is used to check the numeric accuracy of the background solution. We tune the time step such that the relative difference between the two sides of Eq.~\eqref{eq:bg_cons} is less than $10^{-7}$.

To compute the linear perturbations that determines the primordial power spectra, we switch to the conformal time $\tau$. The linear and gauge-invariant scalar perturbation $\mathcal{R}$ (Mukhanov-Sasaki variable) and tensor perturbation $h$ are evolved in Fourier space. The evolution equations are~\citep{Mukhanov92}
\begin{eqnarray}
  \mathcal{R}^{\prime\prime} + 2\frac{z^\prime}{z} \mathcal{R}^\prime + k^2\mathcal{R} &=& 0, \\
  h^{\prime\prime} + 2\mathcal{H} h^\prime + k^2h &=& 0,
\end{eqnarray}
where $z\equiv -a\frac{\dot \phi}{HM_p}$. We apply the adiabatic Bunch-Davies vacuum initial conditions when the Fourier modes are deep inside the horizon ($k\gg \mathcal{H}$),
 \begin{eqnarray}
   \mathcal{R} \rvert_{k\gg\mathcal{H}} & = &  \frac{k^{3/2}}{2\pi z \left(k^2-\frac{z^{\prime\prime}}{z}\right)^{1/4}}e^{-ik\tau}, \\
   h \rvert_{k\gg\mathcal{H} } & =  &  \frac{\sqrt{2}k^{3/2}}{\pi a \left(k^2-\frac{a^{\prime\prime}}{a}\right)^{1/4}}e^{-ik\tau}. 
 \end{eqnarray}
 Here for numeric convenience we have normalized the initial conditions of $\mathcal{R}$ and $h$ such that the dimensionless scalar and tensor power spectra can be evaluated with $\mathcal{P}_S = \lvert\mathcal{R}(k)\rvert^2$ and $\mathcal{P}_T = \lvert h(k)\rvert^2$ when the mode is superhorizon ($k\ll \mathcal{H}$) and both $\mathcal{R}(k)$ and $h(k)$ freeze~\citep{Huang25}. Moreover,  we use analytic solutions in the slow-roll phases to enhance the numeric efficiency\footnote{This is done piecewisely, as the slow-roll parameters are slowly varying.}.

\begin{figure}
  \centering
  \plotone{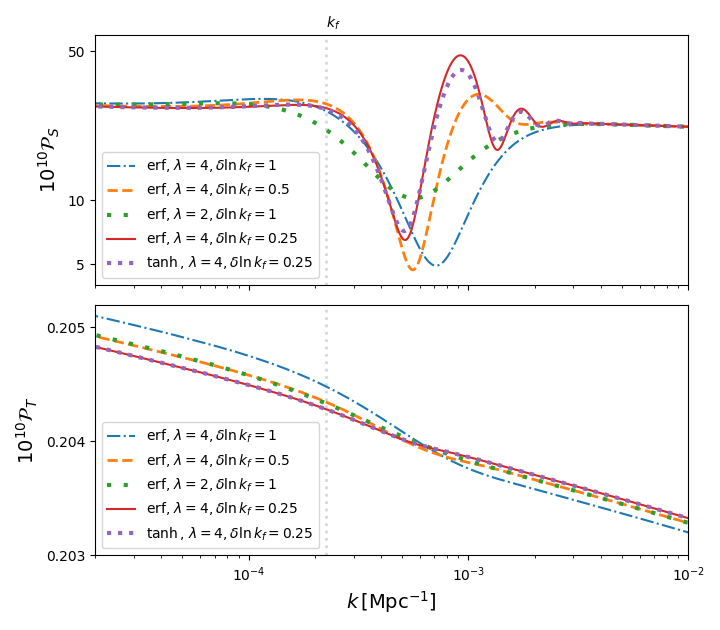}
  \caption{Primordial scalar (upper panel) and tensor (lower panel) power spectra \label{fig:Pk}}
\end{figure}

During the transient fast-roll phase, the energy gap in the potential is converted into kinetic energy, causing the phase-space trajectory to deviate from the slow-roll attractor. This deviation typically damps in an oscillatory manner, generating wiggles (oscillatory features) in the primordial scalar power spectrum. Figure~\ref{fig:Pk} shows several representative examples, with the key parameters $\lambda$ and $\delta\ln k$ indicated in the figure legend. Throughout this work, unless otherwise stated, parameters not explicitly shown in figures assume the default values listed in the last column of Table~\ref{tab:params}.
 From these examples, we observe that the amplitude and width of the main suppression feature in the scalar power spectrum are approximately proportional to $\lambda$ and $\delta\ln k$, respectively. In contrast, the tensor power spectrum exhibits a relatively simple step-like feature that  resembles the form of the inflaton potential. The amplitude and width of this tensor step are controlled by $\lambda\delta\ln k$ and $\delta\ln k$, respectively. The three-parameter model successfully captures the main suppression behavior and the subdominant oscillatory patterns in the scalar power spectrum, while also maintaining consistency between the scalar and tensor power spectra within the single-field inflation framework. In comparison, the phenomenological parameterization adopted in J25—where the scalar power spectrum is suppressed by a factor of $\left(1-e^{-\sqrt{k/k_f}}\right)$ and the tensor power spectrum is neglected—does not account for these finer details.
 
MCMC simulations were performed using cobaya and CAMB~\citep{COBAYA, COSMOMC, CAMB}, in which the default power-law primordial power spectrum was replaced with numerical results from the $\mathrm{erf}$ and $\tanh$ models. The analysis incorporates the following CMB datasets: Planck 2018 PR3 low-$\ell$ EE and TT~\citep{Planck18Params}, joint SPT+ACT+Planck likelihood at high-$\ell$~\citep{SPT25, ACTCmbOnly}, ACT DR6 lensing measurements~\citep{ACTDR6lensing}, and Bicep-Keck 2018 BB power spectrum data~\citep{BicepKeck2018}. For BAO we use the DESI DR2 data~\citep{DESIDR2}. Chain convergence was assessed using the Gelman–Rubin statistic and the Geweke diagnostic.

The results of parameter inference are summarized in Table~\ref{tab:res}. We find little correlation between the step-feature parameters ($\lambda$, $\ln\frac{k_f}{k_p}$, $\delta\ln k_f$) and other cosmological parameters. As a consequence, the posteriors of $\tau$, $\Omega_m$ and $hr_d$ are similar to those in the six-parameter $\Lambda$CDM case that have been reported in~\cite{SPT25}. 
\begin{table*}
  \caption{Constraints on cosmological parameters (mean and 68\% limit) \label{tab:res}}
  \begin{tabular}{lllll}
    \hline
    \hline
    parameter & CMB, no-step & CMB, $\mathrm{erf}$ & CMB, $\tanh$  & CMB-no-lowP+BAO, no-step\\
    \hline
    $\Omega_b h^2$ & $0.02242\pm 0.00010 $ & $0.02243\pm 0.00010$ &  $0.02243\pm 0.00010$ &  $0.02254 \pm 0.00009$ \\
    $\Omega_c h^2$ & $0.1198\pm 0.0012$ & $0.1200\pm 0.0012$ & $0.1199\pm 0.0012$ & $0.1170\pm 0.0007$  \\
    $100\theta_{\rm MC}$ & $1.04074\pm 0.00024$ & $1.04073\pm 0.00024$ & $1.04074\pm 0.00024$ & $1.04096\pm 0.00022$ \\
    $\tau$ & $0.0557^{0.0071}_{-0.0081}$ & $0.0568\pm 0.0081$  & $0.0566\pm 0.0079$ & $0.104\pm 0.017$ \\
    $\ln\left[10^{10}A_s\right]$ & $3.050\pm 0.013$ & $3.052\pm 0.013$ & $3.052\pm 0.013$ & $3.121\pm 0.024$ \\
    $n_s$ & $0.9696\pm 0.0037$ & $0.9685\pm 0.0038$ & $0.9685\pm 0.0038$ & $0.9768\pm 0.0032$ \\
    $r$ (95\% limit) & $<0.0342$ & $<0.0360$ & $<0.0354$ & $<0.0353$ \\
    $\lambda$  (95\% limit) & - & $<6.2$ & $<5.5$ & - \\
    $\ln\frac{k_f}{k_p}$ & - & unconstrained & unconstrained & - \\
    $\delta\ln k_f$ & - & unconstrained & unconstrained & - \\
    \hline
    $\Omega_m$ & $0.3146\pm 0.0070$ & $0.3156\pm 0.0070$ & $0.3153\pm 0.0070$ &  $0.2985\pm 0.0037$ \\
    $\frac{r_d}{\mathrm{Mpc}}$ & $147.10\pm 0.30$ & $147.05\pm 0.30$  & $147.06\pm 0.30$ & $147.70\pm 0.19$ \\
    $\frac{H_0}{\mathrm{km/s/Mpc}}$ & $67.40\pm 0.48$ & $67.33\pm 0.48 $  & $ 67.36\pm 0.49$ & $68.54\pm 0.28$ \\
    $S_8$ & $0.833 \pm 0.011$ & $0.835\pm 0.011$  & $0.835\pm 0.012$ & $0.833\pm 0.010$  \\
    best-fit $\chi^2_{\rm CMB}$ & $1517.3$ & $1512.5$ & $1511.9$ & - \\
    AIC & $1531.3$ & $1532.5$ & $1531.9$ & - \\    
    \hline
  \end{tabular}
\end{table*}

    The comparison of the no-step, $\mathrm{erf}$, and $\tanh$ models based on the best-fit $\chi^2_{\rm CMB}$ and Akaike Information Criterion (AIC) is summarized at the bottom of Table~\ref{tab:res}. The AIC is defined as:
    \begin{equation}
    \mathrm{AIC} = \chi^2_{\min} + 2n_p,
    \end{equation}
    where $\chi^2_{\min}$ is the minimum chi-square and $n_p$ is the number of free parameters. A lower AIC indicates a better balance between goodness of fit and model complexity, with the penalty term $2n_p$ serving to prevent overparameterization. We find that the differences in AIC among the three models are small ($\Delta\mathrm{AIC}\lesssim 1$), suggesting no statistically significant preference for any one model over the others.

\begin{figure}
  \centering
  \plotone{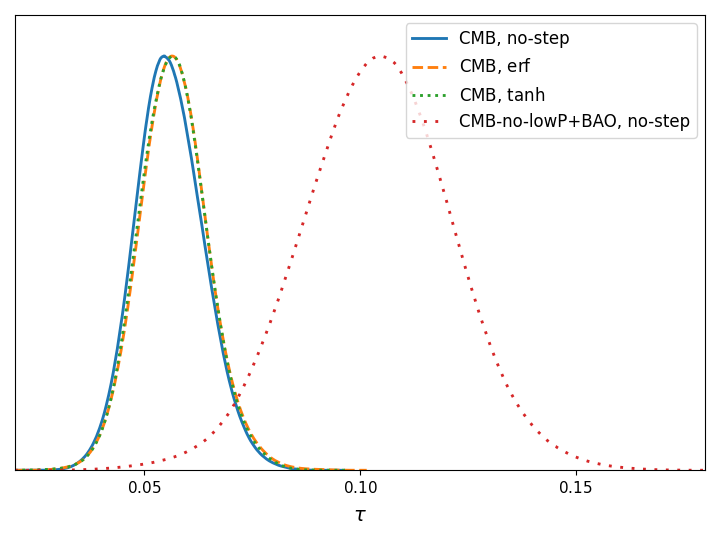}
  \caption{Marginalized constraint on $\tau$. \label{fig:tau}}
\end{figure}

\begin{figure}
  \centering
  \plotone{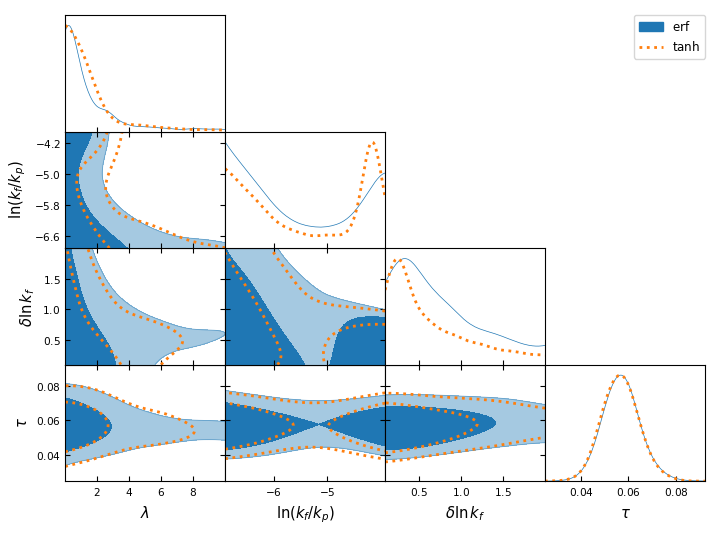}
  \caption{Marginalized $68\%$ and $95\%$ constraints on $\tau$ and step-feature parameters.\label{fig:step}}
\end{figure}

Figure ~\ref{fig:tau} presents the marginalized constraints on $\tau$ derived from various combinations of datasets and models. For both the $\mathrm{erf}$ and $\tanh$ models, the introduction of a step-like feature in the inflaton potential increases $\tau$ by only a few percent. The $\sim 2.5\sigma$ tension in $\tau$ between the full CMB data and the CMB-no-lowP+BAO combination persists. To examine whether this negative result is sensitive to prior choices, Figure~\ref{fig:step} shows the triangle plot for step-feature parameters and $\tau$. No sub-region of the parameter space exhibits a preference for higher $\tau$ values. 

\begin{figure}
  \centering
  \plotone{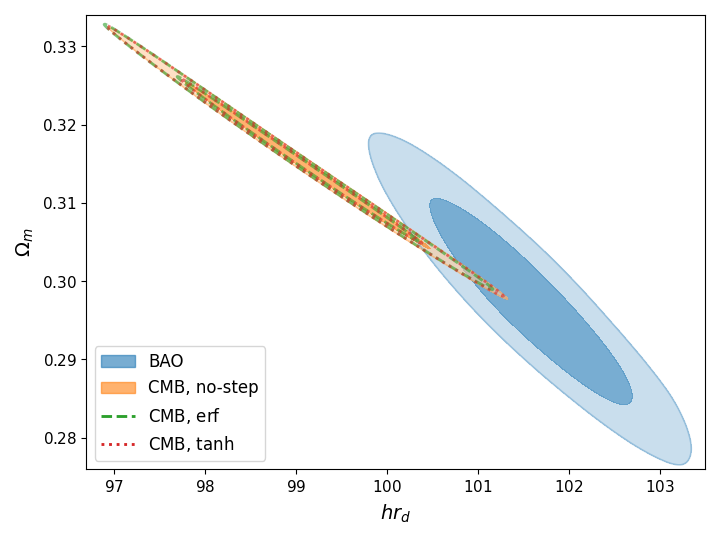}
  \caption{Marginalized 68\%CL and 95\%CL constraints on $r_dh$ and $\Omega_m$. \label{fig:rdomm}}
\end{figure}
Figure~\ref{fig:rdomm} visualizes the CMB-BAO tension in $\Omega_m$-$hr_d$ space. The tension in terms of number of sigmas ($n_\sigma$) can be computed with Gaussian approximation~\citep{Huang24}, 
\begin{equation}
  n_\sigma = \sqrt{2}\,\mathrm{erfc}^{-1}\left(e^{-\chi^2/2}\right),
\end{equation}
where $\mathrm{erfc}^{-1}$ is the inverse of complementary error function and
\begin{equation}
  \chi^2 = (\mathbf{v}_{\rm CMB} - \mathbf{v}_{\rm BAO})^T\left(\mathrm{Cov}_{\rm CMB}+\mathrm{Cov}_{\rm BAO}\right)^{-1}(\mathbf{v}_{\rm CMB} - \mathbf{v}_{\rm BAO}).
\end{equation}
Here the vector $\mathbf{v}$ and matrix $\mathrm{Cov}$ are the mean and covariance matrix of $\begin{pmatrix}\Omega_m  \\ hr_d\end{pmatrix}$, obtained from the Markov chains.
For no-step, $\mathrm{erf}$ and $\tanh$ models, we find $n_\sigma = 2.0$, $2.2$, and $2.1$, respectively.

\section{Discussion and Conclusions } \label{sec:conc}

This work extends the study of J25 by implementing a physical inflationary model that suppresses large-scale primordial curvature perturbations. Unlike the single-parameter phenomenological model in J25 where only the location of feature is specified, our physical framework introduces three additional parameters ($r$, $\lambda$, and $\delta\ln k_f$), enabling a richer set of features, including oscillations in the power spectrum. Nevertheless, we find that punctuated inflation—driven by a step-like feature in the inflaton potential—has a negligible effect on the inferred $\tau$ and does not resolve the CMB-BAO tension. This null result is initially surprising, since a suppression of primordial power could, in principle, compensate for the $\tau$-induced polarization signal. Our findings thus raise a key question: what physical mechanism prohibits a higher $\tau$ value in this model?

\begin{figure}
  \centering
  \plotone{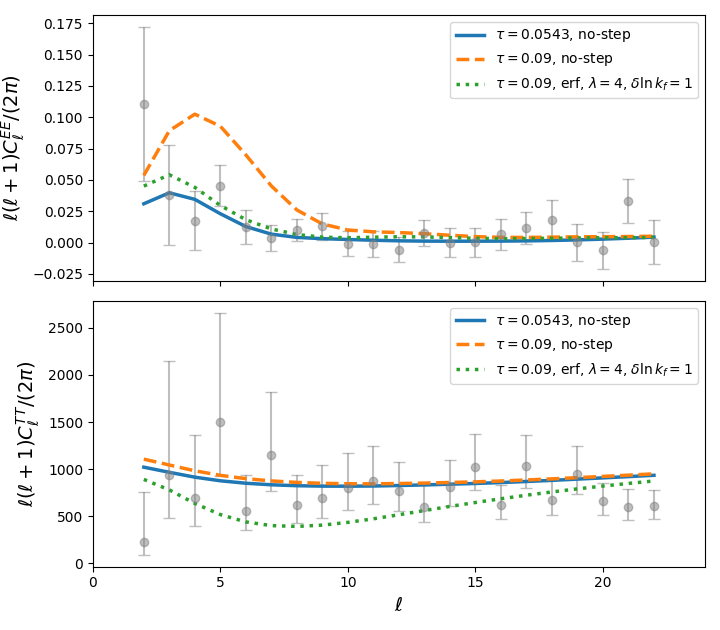}
  \caption{Theoretical CMB EE (upper panel) and TT (lower panel) power spectra, and Planck 2018 data points.  \label{fig:Cl}}
\end{figure}

As Figure~\ref{fig:Cl} shows, a properly chosen step-like feature in the inflaton potential compensates for the effect of a higher optical depth $\tau$ on the theoretical CMB EE power spectrum. However, the specific feature required to raise $\tau$ to approximately $0.09$ suppresses the CMB TT power spectrum excessively, resulting in tension with the Planck temperature data. Thus, joint analyses with the full CMB data would not allow a large step-like feature that substantially raises $\tau$.
The qualitative discussion above applies to a broader range of models that suppresses the large-scale primordial power. We may expect that in general it is difficult to substantially raise $\tau$ via alternative inflation models.

Within the $\Lambda$CDM framework, there is another possibility of raising $\tau$ with non-standard reionization models. This possibility has been discussed in details in J25 and \cite{Obied18}. However, recent direct astrophysical observation by James Webb Space Telescope seems to support a rapid late-time reionization which agrees with a low $\tau\lesssim 0.06$~\citep{Elbers25b, Munoz24}. 

\section*{Acknowledgements}

This work  is supported by National SKA Program of China No. 2020SKA0110402, the National Natural Science Foundation of China (NSFC) under its Key Program (Grant No. 12533002) and General Program (Grant No. 12073088), and National key R\&D Program of China (Grant No. 2020YFC2201600).

\section*{Data Availability}

The data (source codes and Markov chains) underlying this article are available at \url{http://zhiqihuang.top/codes/pinpow.tar.gz}.


\bsp	
\label{lastpage}
\end{document}